\documentclass{webofc}
\usepackage[varg]{txfonts}   % Web of Conferences font
\newcommand{\eps}{\epsilon}
\begin{document}
\title{The $T_{c\bar{s}}(2900)$ as a threshold effect from the interaction of the $D^*K^*$, $D^*_s\rho$ channels}
%
% subtitle is optionnal
%
%%%\subtitle{Do you have a subtitle?\\ If so, write it here}

\author{\firstname{Raquel} \lastname{Molina}\inst{1}\fnsep\thanks{\email{Raquel.Molina@ific.uv.es}} \and
        \firstname{Eulogio} \lastname{Oset}\inst{1}\fnsep\thanks{\email{Eulogio.Oset@ific.uv.es}}
        % etc.
}

\institute{Departamento de F\'{\i}sica Te\'orica and IFIC,
Centro Mixto Universidad de Valencia-CSIC,
Institutos de Investigaci\'on de Paterna, Aptdo. 22085, 46071 Valencia, Spain
          }

\abstract{%
 We investigate the $D^*K^*$ and $D^*_s\rho$ interaction in coupled channels within the hidden gauge formalism. A structure is developed around their thresholds, short of producing a bound state, which leads to a peak in the $D_s^+ \pi^-$ mass distribution in the $B^0 \to \bar{D}^0 D_s^+ \pi^-$ decay compatible with the experimental data. We conclude that the interaction between the $D^*K^*$ and $D^*_s\rho$ is essential to produce the cusp structure that we associate to the recently seen $T_{c\bar{s}}(2900)$, and that its experimental width is mainly due to the decay width of the $\rho$ meson. The peak obtained together with a smooth background reproduces fairly well the experimental mass distribution observed in the $B_0 \to \bar{D}^0 D_s^+ \pi^-$ decay. 
}
\maketitle
\section{Introduction}
\label{intro}
The $D^*K^*$ system was investigated in \cite{branz} and three states were found corresponding to $I=0;J^P=0^+,1^+$ and $2^+$. The $2^+$ state was identified with the $D^*_{s2}(2573)$ state, and served to set the scale for the regularization of the loops, allowing predictions in the other sectors. There, the $I=1$ interaction of the $D^*K^*$ and $D^*_s\rho$ channels was also studied and, a cusp was found for $J=0$ and $J=1$ around the $D^*_s\rho$ threshold.  

Recently, the LHCb Collaboration has observed an state in the $D^+_s\pi^-$, $D^+_s\pi^+$ mass distributions in the $B^0\to \bar{D}^0D^+_s\pi^-$ and $B^+\to D^-D_s^+\pi^+$ decays, respectively, at $2900$~MeV \cite{lhcb1,lhcb2}. Indeed, the state branded as $T_{c\bar{s}}(2900)$ with $J^P=0^+$, as seen in $D^+_s\pi^-$ and $D^+_s\pi^+$,  exhibits an $I=1$ character and it has also been associated with $J^P=0^+$. On the other hand, $2900$~MeV is just the threshold of the $D^*K^*$ channel. Thus, one is finding a $I=1$ $J^P=0^+$ state in the threshold of $D^*K^*$ (the $D^*_s\rho$ is only $14$ MeV below neglecting the $\rho$ width), which could correspond to the cusp found in \cite{branz}.

In the present work we look again at the interaction of $D^*K^*$ and $D^*_s\rho$ channels, taking into account the $K^*$ and $\rho$ widths and also the decay of the states found into the $D_s\pi$ channel where it has been observed, comparing our results with the recent experimental findings. 

\section{Formalism}
\begin{figure}
 \begin{center}
  \begin{tabular}{cc}
  \includegraphics[scale=0.5]{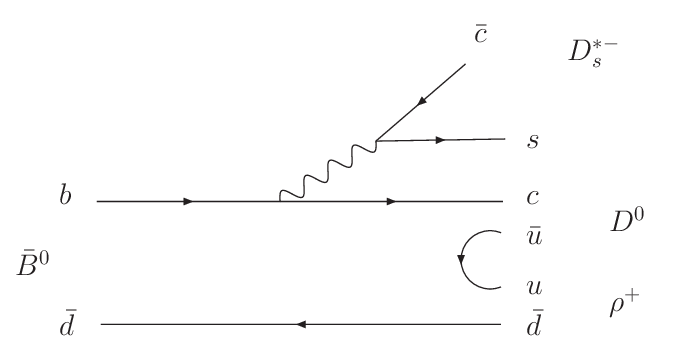}&
  \includegraphics[scale=0.5]{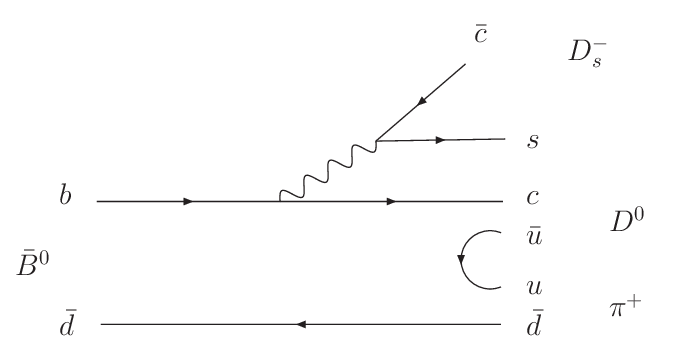}
  \end{tabular}
  \end{center}
  \caption{Left: $\bar{B}^0$ decay to $D^{*-}_sc\bar{d}$ with hadronization of the $c\bar{d}$ pair to produce $D^{*-}_sD^0\rho^+$. Right: $\bar{B}^0$ decay into $D_s^-D^0\pi^+$ (contribution to the background).}
  \label{fig:1}
 \end{figure}

  In Ref.~\cite{lhcb1} a peak is found in the $D_s\pi$ invariant mass in the $B^0\to\bar{D}^0D^+_s\pi^-$ and $B^+\to D^-D^+_s\pi^+$ decays. In order to have a $b$ quark rather than a $\bar{b}$ quark, we look at the reaction $\bar{B}^0\to D^0D_s^{*-}\rho^+$. We produce this state with the external emission Cabibbo favored decay shown in Fig.~\ref{fig:1} (left). In Fig.~\ref{fig:1} (right) we depict the direct decay $\bar{B}^0\to D_s^-D^0\pi^+$ considered as background.
 
\begin{figure}
  \centering
  \includegraphics[scale=0.5]{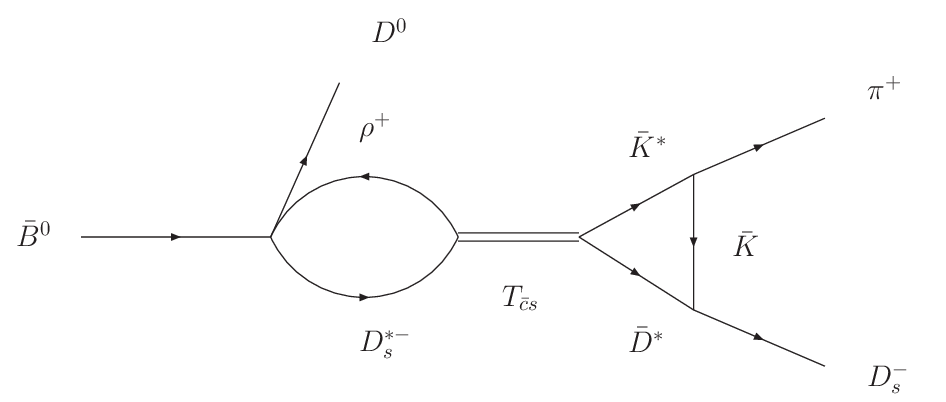}
  \caption{Mechanism by means of which the resonance is produced and decays into $\pi^+D_s^-$.}
  \label{fig:4}
 \end{figure}

\begin{figure}
  \centering
  \includegraphics[scale=0.6]{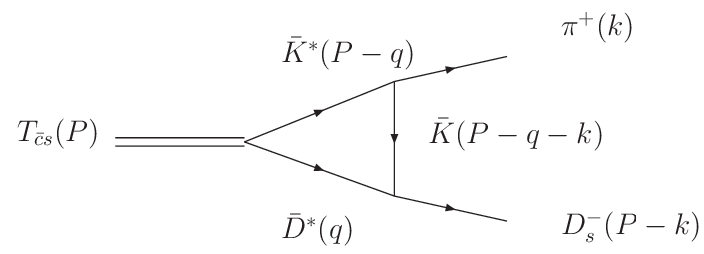}
  \caption{Triangle diagram accounting for the $R\to\pi \bar{D}_s$ decay of the $R$ resonance of $I=1$ generated with the $\rho\bar{D}_s$, $\bar{D}^*\bar{K}^*$ coupled channels.}
  \label{fig:5}
 \end{figure}

We evaluate the scattering matrix using the Bethe-Salpeter equation in the $D^*K^*$ and $D^*_s\rho$ channels,
 \begin{equation}
  T=[1-VG]^{-1}V\ ,\label{eq:bethe}
 \end{equation}
 with $G$ the diagonal loop function for the intermediate mesons and $V$ the transition potential. However, the state is observed in $D_s\pi$, hence, the mechanism by means of which the reaction proceeds is given in Fig.~\ref{fig:4}.
 The amplitude for the process of Fig.~\ref{fig:4} is given by,
 \begin{equation}
  t=aG_{\rho D^{*}_s}(M_\mathrm{inv})t_{\rho D^*_s,K^*D^*}(M_\mathrm{inv})\tilde{V}(\pi D_s, M_\mathrm{inv})\label{eq:t}
 \end{equation}
 where $a$ is a normalization constant that we do not evaluate, unnecessary to show the shape of the $\pi D_s$ mass distribution in the $\bar{B}^0$ decay, and $M_\mathrm{inv}$ is the invariant mass distribution of the $D_s\pi$ final state. The vertex function $\tilde{V}$ corresponding to the triangle loop of Fig.~\ref{fig:5} can be easily evaluated. 
 
  We assume the resonance to be in $J=0$, and that the vectors have small momenta with respect to their masses. Then, the structure of the triangle diagram of Fig.~\ref{fig:5} is given by
\begin{eqnarray}
 \tilde{V}=-i\int \frac{d^4q}{(2\pi)^4}\eps_{\bar{K}^*}^l\eps_{\bar{D}^*}^l\eps_{\bar{K}^*}^i\eps_{\bar{D}^*}^j\frac{(2k+q)^i(2k+q)^j}{(P-q-k)^2-m^2_K+i\eps}\frac{\theta(q_\mathrm{max}-q)}{(P-q)^2-m_{K^*}^2+i\eps}\frac{1}{q^2-m^2_{D^*}+i\eps}\ .\label{eq:vt}
\end{eqnarray}
The loop function $\tilde{V}$ is naturally regularized with a cutoff $q_\mathrm{max}$, the same one used to regularize the $D^*K^*$ and $D^*_s\rho$ loops when studying their interactions. The equivalent $q_\mathrm{max}$ used in \cite{branz} was $1100$~MeV. We find,

\begin{eqnarray}
 &&\tilde{V}=-\int\frac{d^3q}{(2\pi)^3}\frac{(2\vec{k}+\vec{q})^2}{8\omega_{K^*}(q)\omega_{D^*}(q)\omega_K(\vec{q}+\vec{k})}\frac{1}{P^0-\omega_{D^*_s}(q)-\omega_{K^*}(q)+i\eps}\nonumber\\&&\times\left\{\frac{1}{P^0-k^0-\omega_{D^*}(q)-\omega_K(\vec{q}+\vec{k})+i\eps)}\right.\left.+\frac{1}{k^0-\omega_{K^*}(q)-\omega_K(\vec{q}+\vec{k})+i\eps}\right\}\ ,\label{eq:tl}
\end{eqnarray}
which shows the different cuts of the loop diagram when pairs of the internal particles of the loop are placed on-shell.

Then, we consider that the transition amplitude for $\bar{B}^0\to D^0D^-_s\pi^+$ is given by a constant background (considering the dominance of s-wave in the coupling of the bottom meson to the pseudoscalars), see Fig.~\ref{fig:1} (right), together with the scattering amplitude of the diagram in Fig.~\ref{fig:4}, which accounts for the interaction of the $VV$ coupled channels. It reads as
\begin{equation}
  t'=aG_{\rho D^*_s}(M_\mathrm{inv})t_{\rho D^*_s,K^*D^*}(M_\mathrm{inv})\tilde{V}(\pi D_s, M_\mathrm{inv})+b\label{eq:t1}
 \end{equation}

Therefore, the mass distribution of $\pi D_s^-$ in the $\bar{B}^0$ decay is given by,
\begin{equation}
 \frac{d\Gamma}{dM_{\mathrm{inv}}}=\frac{1}{(2\pi)^3}\frac{1}{4M^2_B}p_{D^0}\tilde{p}_\pi\vert t'\vert^2\ ,\label{eq:dg}
\end{equation}
where $p_{D^0}=\frac{\lambda^{1/2}(M^2_B,m^2_{D^0},M^2_{\mathrm{inv}})}{2M_B}$ and $\tilde{p}=\frac{\lambda^{1/2}(M^2_\mathrm{inv},m^2_{D_s},m^2_\pi)}{2M_\mathrm{inv}}$.  
 
\subsection{Results}
\begin{figure}
\centering
 \includegraphics[scale=0.3]{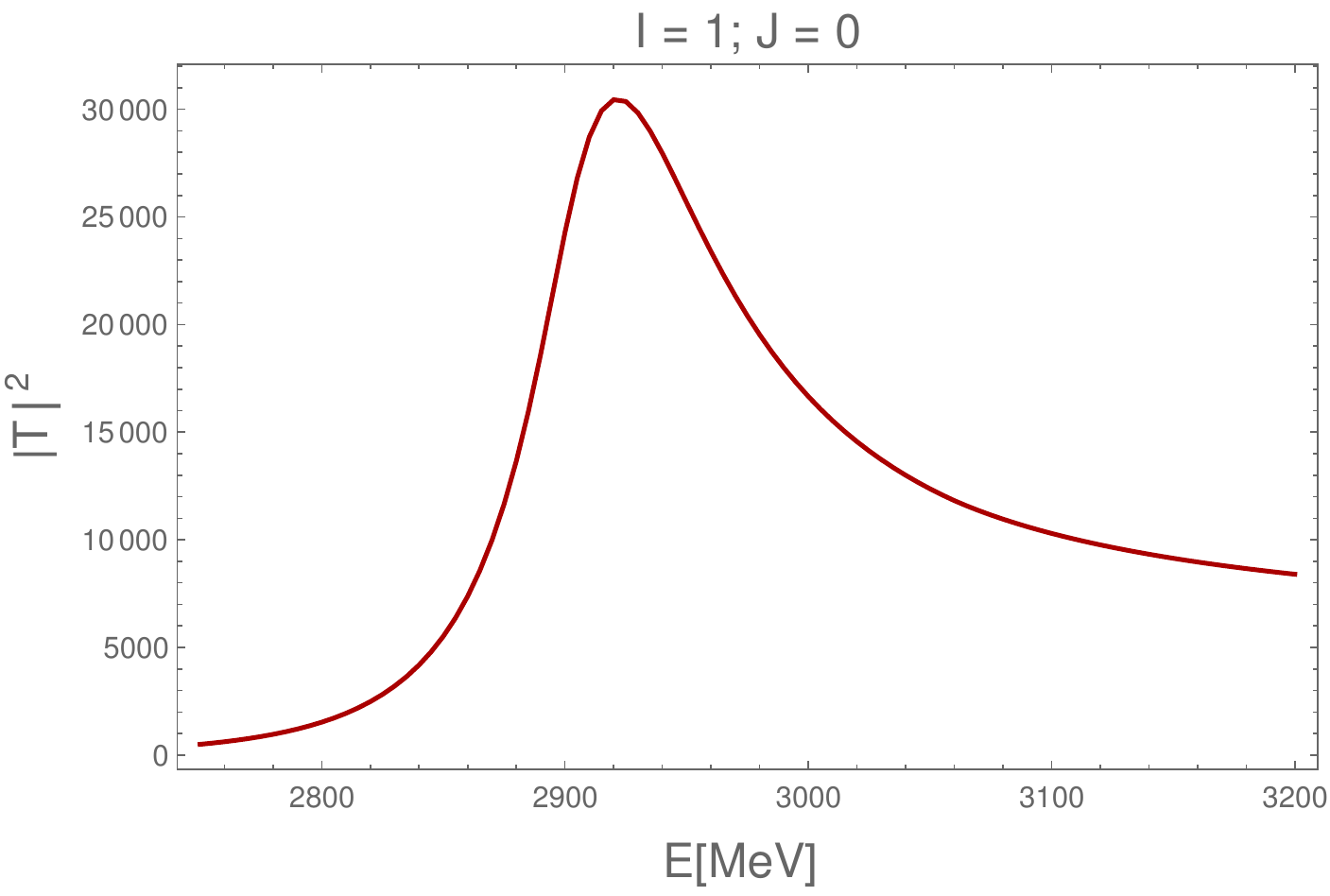}
 \caption{$\vert T\vert^2$ for $C=1;S=1;I=1;J=0$ with $\alpha=-1.474$.}.
 \label{fig:t101}
\end{figure}

We take into account the decay widths of the vector mesons $K^*$ and $\rho$ by means the convolution of the $G$ function in Eq.~(\ref{eq:bethe})~\cite{tcsbar}. The result for the $T$ matrix in $I=1;J=0$ is shown in Fig.~\ref{fig:t101}. The cusp obtained for $J=0$ has become wider. The position of the cusp is similar, it shows up slightly above the $D^*K^*$ threshold and around $2920$ MeV, with a width coming basically from the decay of the $\rho$ into $\pi\pi$. We have also obtained visible peaks in the scattering amplitudes for $J=1$ and $2$~\cite{tcsbar}.

\begin{figure}
 \centering
 \begin{tabular}{cc}
 \hspace{-0.5cm}\includegraphics[scale=0.4]{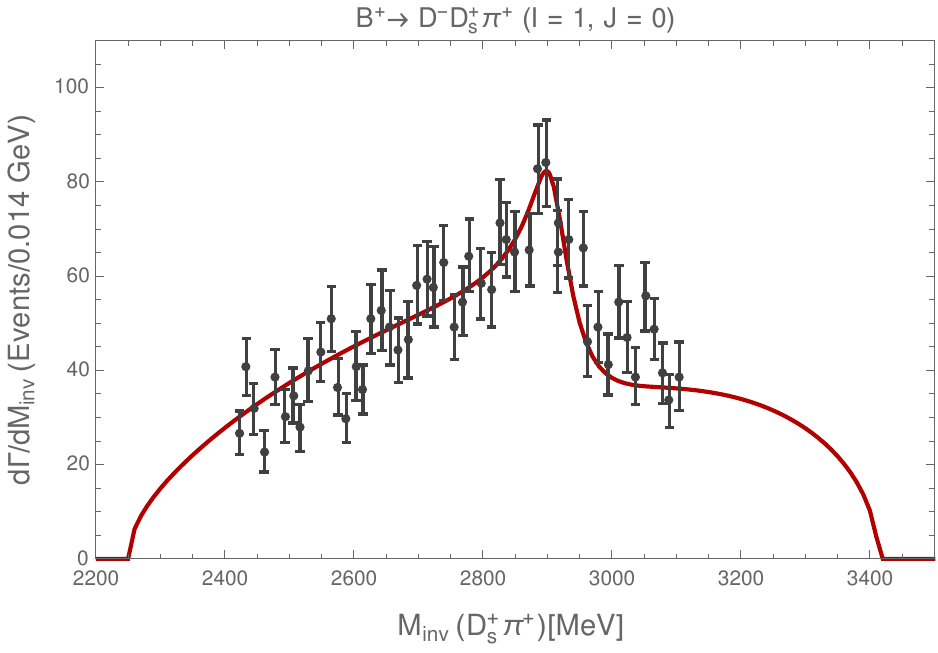}&\includegraphics[scale=0.41]{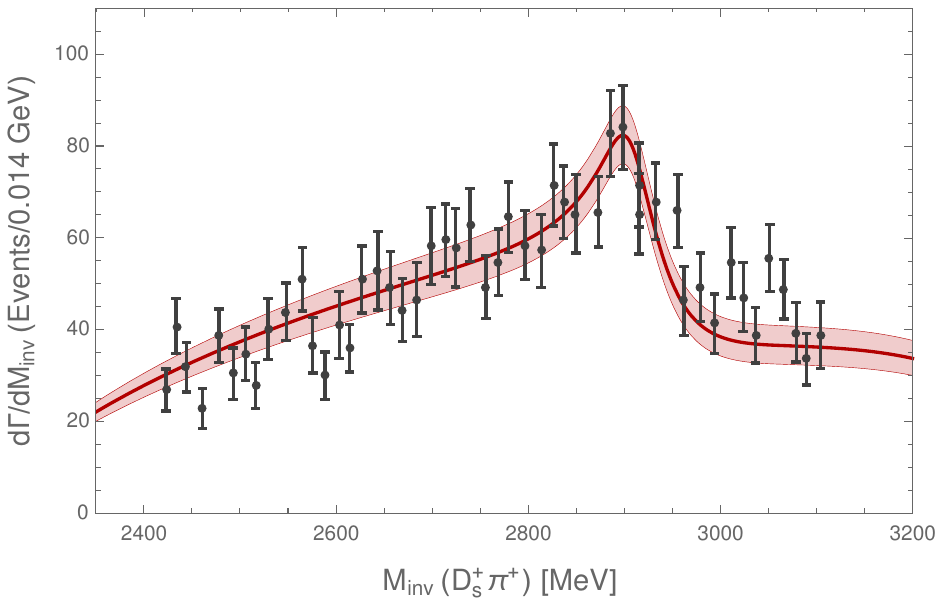}\end{tabular}
 \caption{Left: invariant mass distribution for $D_s\pi$ from the decay $B\to \bar{D}D_s\pi$ compared to the experimental data from Ref.~\cite{lhcb1,lhcb2}. Right: the same but with the error band obtained by changing the parameter for the background ($b$) 5\% up and down.}
 \label{fig:compexp}
\end{figure}

Finally, we show the result of the invariant mass distribution of the decay $\bar{B}^0\to D^-_s D^0\pi^+$, Eq.~(\ref{eq:dg}), in comparison with the LHCb experimental data \cite{lhcb1,lhcb2} in Fig.~\ref{fig:compexp} (left). In Eq.~(\ref{eq:dg}), we adjusted the constants $a$ and $b$ to reproduce well the experimental data around the $T_{c\bar{s}}(2900)$ resonance, and we obtain $a=2.1\times 10^3$ and $b=-1.45\times 10^3$. As can be seen, our model describes well the experimental data. A peak is obtained around the threshold of the $D^*K^*$ channel. Since these results were obtained fixing the subtraction constant to obtain the $T_{cs}(2900)$, this also supports the molecular picture of this state as $D^*\bar{K}^*$ of \cite{raquel}. Thus, our model strongly supports the $T_{c\bar{s}}(2900)$ as a cusp structure originated by the non-diagonal interaction $D^*K^*\to D^*_s\rho$, with a width mainly due to the decay of the $\rho$ meson into $\pi\pi$

Finally, it is interesting to give a band of errors by changing the background, we do this to show the sensitivity of the results to this background. We have done this, keeping the value of $a$, needed to get the strength of the peak of the distribution, by varying the parameter $b$ of the background by $5\%$ (up and down). This is shown in Fig.~\ref{fig:compexp} (right). The band obtained overlaps with the errors of the data.

\bibliography{bibliotc}
\end{document}